\documentclass{JFM-FLM_Au}

\usepackage{graphicx}
\usepackage{booktabs}
\usepackage{amsmath}
\newcommand{\ycp}{y_{cp}^+}
\newcommand{\lamcp}{\lambda_{z,cp}^+}
\newcommand{\TmTw}{T_m/T_w}
\newcommand{\Retcp}{\Rey_{\tau,cp}}
\newcommand{\ncases}{twelve}
\newcommand{\jfmnb}[1]{JFM Notebook: \href{#1}{\nolinkurl{#1}}}
\newcommand{\condkey}{$\TmTw=1.1$ and $T_w=277$\,K (orange right triangles), $\TmTw=1.2$ and $T_w=277$\,K (light blue downward triangles), $\TmTw=1.3$ and $T_w=277$\,K (green upward triangles), $\TmTw=1.1$ and $T_w=300$\,K (brown stars), $\TmTw=1.2$ and $T_w=300$\,K (red circles), $\TmTw=1.1$ and $T_w=330$\,K (purple squares), $\TmTw=0.85$ and $T_w=330$\,K (magenta diamonds), $\TmTw=0.9$ and $T_w=360$\,K (teal pentagons) and $\TmTw=0.83$ and $T_w=360$\,K (gold hexagons)}

\lefttitle{D. Modesti and A. Ceci}
\righttitle{Journal of Fluid Mechanics}

\title{Direct numerical simulation of turbulent heat transfer in water with temperature-dependent viscosity and conductivity}

\author{Davide Modesti\aff{1} \and Alessandro Ceci\aff{1}}

\affiliation{\aff{1}Gran Sasso Science Institute, Viale Francesco Crispi 7, 67100 L'Aquila, Italy}

\corresau{Davide Modesti, \email{davide.modesti@gssi.it}}

\begin{document}
\maketitle

\begin{abstract}
We study turbulent forced convection of liquid water in a plane channel by direct numerical simulation (DNS), with temperature-dependent viscosity and thermal conductivity. The simulations cover wall heating and cooling, mean-to-wall temperature ratios from 0.83 to 1.3, wall temperatures from 277 to 360\,K and wall Prandtl numbers from 2 to 12. Compared with constant-property flow at the same bulk Reynolds number, the friction coefficient changes by $-39\,\%$ to $+32\,\%$, the heat transfer coefficient referred to the wall Prandtl number changes by up to a factor of $2.3$, and empirical correlations deviate from the DNS by up to $26\,\%$. Instantaneous fields and spectra show that the wall friction Reynolds number misrepresents the range of turbulent scales, whereas near-wall streaks of velocity and temperature have a common spacing when scaled with the local viscous length. The density is constant, so the mean velocity and temperature defects in the core follow those of constant-property flow. To predict friction and heat transfer, we map the mean profiles onto those of constant-property flow through integral transformations derived from the mean momentum and energy balances, with empirical corrections of the eddy viscosity and diffusivity calibrated on the DNS. This approach, developed for air, is extended to liquids, whose conductivity is not proportional to the viscosity and whose Prandtl number varies across the channel. The inverse transformations predict friction and heat transfer within $2\,\%$ of the DNS, except at the lowest Reynolds numbers. The same corrections, introduced in the Spalart--Allmaras model, halve its friction error.
\end{abstract}


%=====================================================================================================
\section{Introduction}\label{sec:intro}

Water is the working fluid of most heat exchangers, from district heating and power plants to the cooling circuits of electronics, nuclear reactors and fusion devices. In many of these applications the temperature difference between the wall and the fluid is large enough for the molecular properties to change substantially across the boundary layer. The dynamic viscosity of liquid water is particularly sensitive, as it halves between $300$ and $340$\,K and varies by a factor of six between the freezing and the boiling point, whereas the thermal conductivity changes by about $20\,\%$ and the density by less than $5\,\%$. The molecular Prandtl number $\Pran=\mu c_p/\lambda$ can therefore change by a factor of up to six between the wall and the core of a heated or cooled duct, and the classical picture of a passive scalar transported at a single Prandtl number does not apply.

Engineering practice accounts for these effects through empirical corrections of constant-property formulas. For liquids the most common is the factor $(\mu_m/\mu_w)^n$ of \citet{siedertate1936}, applied to the Nusselt number, where $\mu_m$ and $\mu_w$ are the viscosities at the mean and at the wall temperature. \citet{petukhov1970} later proposed different exponents for heating and cooling, together with a separate correction of the friction factor. The exponent of Sieder \& Tate was determined from turbulent data of earlier investigators in which, as the authors note, the viscosity varied little between the wall and the stream, and no range of validity is given. The heat transfer correction of Petukhov was fitted to data with $0.08<\mu_w/\mu_m<40$, whereas his friction correction rests on water data with $0.35<\mu_w/\mu_m<2$, and its accuracy at larger viscosity contrast is uncertain. A predictive theory of friction and heat transfer in liquids with variable properties is not available.

Direct numerical simulation (DNS) and large-eddy simulation of variable-viscosity flows have clarified several aspects of the physics. \citet{zonta2012} simulated water in a channel with one hot and one cold wall and found that the Reynolds shear stress and the friction drop at the hot wall, despite the lower viscosity there. \citet{lee2013,lee2014} studied boundary layers with temperature-dependent viscosity representative of water and reported a drag reduction of $26\,\%$ for a free-stream-to-wall temperature ratio of $0.81$, together with a correction of the temperature law of the wall of \citet{kader1981}. \citet{patel2015,patel2016,patel2017} showed that the semi-local scaling developed for compressible flows \citep{huang1995,trettel2016}, in which the wall distance is scaled with the local viscous length, $\ycp=yu_\tau/\bar\nu$, denoted $y^*$ in the literature on high-speed boundary layers, collapses the mean velocity and the shear stresses of liquid-like and gas-like fluids, though not the turbulence intensities, and extended the analysis to the statistics of a passive scalar. \citet{kaller2019} performed wall-resolved large-eddy simulations of a cooling duct with water and asymmetric heating and documented the modification of the near-wall turbulence at the heated wall. None of these studies, however, addressed the prediction of the friction and heat transfer coefficients over a wide range of conditions.

In a previous study \citep{modesti2024air} we developed a framework for predicting friction and heat transfer in forced air convection with variable properties. Integral transformations derived from the mean momentum and thermal balances map the variable-property profiles onto the constant-property ones, and their inversion, together with accurate constant-property profiles \citep{pirozzoli2023,pirozzoli2024profiles}, returns the friction and heat transfer coefficients within $1$--$2\,\%$ of the DNS. In air the conductivity is proportional to the viscosity and the Prandtl number is constant, while the density changes by a large factor. In liquid water the situation is reversed, since the density varies little, the conductivity is decoupled from the viscosity, and the Prandtl number varies across the layer. Whether the framework carries over to liquids is therefore not obvious.

In this paper we present a DNS database of turbulent channel flow of water with temperature-dependent viscosity and conductivity, covering both wall heating and wall cooling at four wall temperatures, with wall Prandtl numbers between 2 and 12. We first discuss the instantaneous fields and their spectral signature, which show how the variable viscosity changes the scale separation. We then examine the mean profiles and their defect form, extend the transformations to liquids, and use them to predict friction and heat transfer. Finally, since many heat-exchanger computations rely on Reynolds-averaged Navier--Stokes (RANS) models, we show how the same corrections can be introduced in the Spalart--Allmaras model \citep{spalart1994}, using the heat-transfer formulation of \citet{ceci2025}.

%=====================================================================================================
\section{Methodology}\label{sec:method}

\subsection{Governing equations and fluid properties}

We solve the incompressible Navier--Stokes equations at constant density $\rho$ and constant specific heat $c_p$, with temperature-dependent viscosity $\mu(T)$ and conductivity $\lambda(T)$,
\begin{gather}
\frac{\partial u_j}{\partial x_j}=0,\qquad
\rho\left(\frac{\partial u_i}{\partial t}+\frac{\partial u_iu_j}{\partial x_j}\right)=-\frac{\partial p}{\partial x_i}+\frac{\partial}{\partial x_j}\left[\mu\left(\frac{\partial u_i}{\partial x_j}+\frac{\partial u_j}{\partial x_i}\right)\right]+\Pi\,\delta_{i1},\label{eq:ns}\\
\rho c_p\left(\frac{\partial T}{\partial t}+\frac{\partial Tu_j}{\partial x_j}\right)=\frac{\partial}{\partial x_j}\left(\lambda\frac{\partial T}{\partial x_j}\right)+S,\label{eq:energy}
\end{gather}
where $\Pi(t)$ is a spatially uniform pressure gradient and $S(t)$ a spatially uniform heat source, determined at each time step so that the bulk velocity and the mixed mean temperature,
\begin{equation}
u_b=\frac{1}{2h}\int_0^{2h}\bar u\,\mathrm{d}y,\qquad T_m=\frac{1}{2hu_b}\int_0^{2h}\overline{uT}\,\mathrm{d}y,\label{eq:bulk}
\end{equation}
remain constant in time. Here $h$ is the channel half-width, an overline denotes averaging in time and in the homogeneous spatial directions, and a prime denotes fluctuations thereof. The full viscous stress is retained, since the transpose term $\partial_j(\mu\,\partial_iu_j)$ does not vanish when $\mu$ varies. Buoyancy and viscous heating are neglected, as is customary in studies of forced convection. The density is taken as constant, which is consistent with its variation of less than $5\,\%$ over the temperature range considered here. The viscosity follows the Vogel--Fulcher--Tammann law \citep{vogel1921,fulcher1925,tammann1926}, with the constants used for water in the DNS of \citet{zonta2012},
\begin{equation}
\mu(T)=2.414\times10^{-5}\;10^{\,247.8/(T-140)}\ \mathrm{Pa\,s},\label{eq:mu}
\end{equation}
and the conductivity the cubic polynomial
\begin{equation}
\lambda(T)=-0.869+8.94\times10^{-3}\,T-1.584\times10^{-5}\,T^2+7.98\times10^{-9}\,T^3\ \mathrm{W\,m^{-1}\,K^{-1}},\label{eq:lam}
\end{equation}
with $T$ in kelvin and $c_p=4184$\,J\,kg$^{-1}$\,K$^{-1}$. The two laws reproduce the IAPWS formulations \citep{iapws2008viscosity,iapws2011conductivity} to within $1.3\,\%$ and $1.1\,\%$, respectively, between 277 and 365\,K, which is the range of the mean temperature in all cases, and to within $1.7\,\%$ and $1.2\,\%$ between 275 and 383\,K, the range of the instantaneous temperature. Both laws refer to liquid water at atmospheric pressure, and pressure does not enter the equations otherwise, since the density is constant. The dependence of the two properties on pressure is weak, as the IAPWS formulations give changes of less than $0.7\,\%$ up to 5\,MPa and of at most $1.3\,\%$ at 10\,MPa over the same range of temperature, so that the laws apply equally to pressurised liquid water. Each flow case is therefore defined by the bulk Reynolds number, the mean-to-wall temperature ratio $\TmTw$ and the wall temperature $T_w$. The latter fixes the wall Prandtl number $\Pran_w=\mu_wc_p/\lambda_w$, which equals $11.58$, $5.93$, $3.16$ and $2.02$ at $T_w=277$, $300$, $330$ and $360$\,K. Between the wall and the mixed mean temperature, $\mu$ varies by a factor of $1.6$--$4.8$ and $\lambda$ by $4$--$19\,\%$ in the opposite direction, so that the Prandtl number varies by a factor of $1.6$--$5.7$.

The subscripts $w$ and $m$ denote quantities evaluated at the wall temperature and at the mixed mean temperature, respectively. A $+$ superscript denotes normalisation by wall units, namely by the friction velocity $u_\tau=(\tau_w/\rho)^{1/2}$, where $\tau_w=\mu_w\,\mathrm{d}\bar u/\mathrm{d}y|_w$ is the mean wall shear stress, by the viscous length scale $\delta_v=\nu_w/u_\tau$, and by the friction temperature $\theta_\tau=q_w/(\rho c_pu_\tau)$, where $q_w=\lambda_w\,\mathrm{d}\bar T/\mathrm{d}y|_w$ is the mean wall heat flux. Temperatures are measured from the wall value, $\theta=T-T_w$. The bulk and friction Reynolds numbers, the friction coefficient, the Stanton number and the Nusselt number are defined as
\begin{equation}
\Rey_b=\frac{2\rho u_bh}{\mu_m},\quad \Rey_\tau=\frac{u_\tau h}{\nu_w},\quad C_f=\frac{2\tau_w}{\rho u_b^2},\quad St=\frac{q_w}{\rho c_pu_b(T_m-T_w)},\quad Nu=St\,\Rey_b\Pran_m.\label{eq:coef}
\end{equation}
Following \citet{modesti2024air}, we refer to $\TmTw<1$ as wall heating and to $\TmTw>1$ as wall cooling. Contrary to gases, the viscosity of water decreases with temperature, hence for wall cooling the viscosity is highest at the wall.

\subsection{Numerical method and database}

The equations are solved with the GPU flow solver THETA (Turbulent Heat Exchange and Transport Algorithm), which stems from the incompressible solver of \citet{orlandi2000}, used for the channel DNS of \citet{bernardini2014}. THETA uses second-order finite differences on a staggered grid \citep{harlow1965} with the convective terms in divergence form, the three-stage Runge--Kutta scheme of \citet{wray1990} with implicit treatment of the wall-normal viscous terms, and a fractional-step projection in which the Poisson equation is solved by Fourier transforms in the homogeneous directions and a tridiagonal solver in the wall-normal one. At the isothermal walls the viscous and conductive fluxes are evaluated with the wall properties $\mu_w$ and $\lambda_w$. The computational box is $L_x\times L_y\times L_z=6\upi h\times2h\times2\upi h$, with periodicity in $x$ and $z$, no-slip isothermal walls and a wall-normal mesh clustered near the walls according to the natural stretching of \citet{pirozzoli2021grid}.

\begin{table}
\begin{center}
\def~{\hphantom{0}}
\setlength{\tabcolsep}{1.5pt}
\footnotesize
\begin{tabular}{lrrrrrrrrrrrrrrr}
\toprule
Case & $\Rey_b$ & $\Rey_\tau$ & $\Retcp$ & $\TmTw$ & $T_w$ (K) & $\Pran_w$ & $\Delta x^+$ & $\Delta y_w^+$ & $\Delta z^+$ & $\Delta x/\eta_B$ & $\Delta y/\eta_B$ & $\Delta z/\eta_B$ & $10^3C_f$ & $10^3St$ & $Nu$\\
\midrule
L12-B & 23\,201 & 260 & 711 & 1.20 & 300 & 5.93 & 2.4 & 0.69 & 1.6 & 4.9 & 1.4 & 3.3 & 7.00 & 1.836 & 86.0 \\
L12-A & 15\,002 & 144 & 493 & 1.20 & 277 & 11.58 & 1.3 & 0.38 & 0.9 & 4.1 & 1.2 & 2.7 & 8.10 & 1.541 & 70.2 \\
L13 & 19\,463 & 128 & 643 & 1.30 & 277 & 11.58 & 1.2 & 0.34 & 0.8 & 4.2 & 1.2 & 2.8 & 8.02 & 1.792 & 70.4 \\
L083 & 5071 & 353 & 125 & 0.83 & 360 & 2.02 & 3.2 & 1.05 & 2.2 & 2.7 & 0.9 & 1.8 & 5.53 & 1.388 & 41.7 \\
L11-A & 9919 & 158 & 325 & 1.10 & 277 & 11.58 & 1.5 & 0.42 & 1.0 & 3.8 & 1.1 & 2.5 & 8.27 & 1.262 & 66.2 \\
L11-C & 16\,353 & 310 & 493 & 1.10 & 330 & 3.16 & 2.9 & 0.83 & 1.9 & 3.6 & 1.1 & 2.4 & 6.98 & 2.112 & 67.2 \\
H12-B & 49\,341 & 501 & 1370 & 1.20 & 300 & 5.93 & 2.3 & 0.63 & 1.5 & 4.7 & 1.3 & 3.2 & 5.76 & 1.583 & 157.8 \\
H13 & 44\,819 & 262 & 1317 & 1.30 & 277 & 11.58 & 1.2 & 0.33 & 0.8 & 4.3 & 1.1 & 2.8 & 6.34 & 1.506 & 136.1 \\
H083 & 14\,006 & 891 & 317 & 0.83 & 360 & 2.02 & 4.1 & 1.11 & 2.7 & 3.6 & 1.0 & 2.4 & 4.63 & 1.300 & 107.9 \\
H11-B & 28\,541 & 450 & 806 & 1.10 & 300 & 5.93 & 2.1 & 0.56 & 1.4 & 3.8 & 1.0 & 2.5 & 6.13 & 1.442 & 130.2 \\
H09 & 26\,947 & 1082 & 630 & 0.90 & 360 & 2.02 & 5.0 & 1.35 & 3.3 & 4.5 & 1.2 & 3.0 & 4.67 & 1.514 & 144.0 \\
H085 & 12\,507 & 875 & 289 & 0.85 & 330 & 3.16 & 4.0 & 1.09 & 2.7 & 4.4 & 1.2 & 2.9 & 4.72 & 0.965 & 124.4 \\
\bottomrule
\end{tabular}
\caption{Parameters of the DNS database. $\Rey_b=2\rho u_bh/\mu_m$ is the bulk Reynolds number, $\Rey_\tau=u_\tau h/\nu_w$ the friction Reynolds number, $\Retcp$ the equivalent constant-property friction Reynolds number defined in \eqref{eq:retcp}, $T_m$ and $T_w$ the mixed mean and wall temperatures, and $\Pran_w$ the wall Prandtl number. The L cases are computed on a mesh of $2048\times560\times1024$ cells and the H cases on a mesh of $4096\times1120\times2048$ cells, in the streamwise, wall-normal and spanwise directions. $\Delta x^+$, $\Delta z^+$ and $\Delta y^+_w$ are the streamwise, spanwise and wall-normal (at the wall) mesh spacings in wall units, and $\Delta x/\eta_B$, $\Delta y/\eta_B$ and $\Delta z/\eta_B$ the maxima across the channel of the spacings over the local Batchelor scale. $C_f$, $St$ and $Nu$ are the friction coefficient, the Stanton number and the Nusselt number defined in \eqref{eq:coef}.}
\label{tab:flow}
\end{center}
\end{table}

The database (table~\ref{tab:flow}) consists of \ncases{} flow cases, computed on two meshes. The flow cases are denoted with the letters L and H depending on whether the Reynolds number is comparatively low or high, followed by the value of $\TmTw$ (e.g.\ L12 for $\TmTw=1.2$). The ratios $\TmTw=1.1$ and $1.2$ were simulated at more than one wall temperature, and for these the suffixes A, B and C denote $T_w=277$, $300$ and $330$\,K, respectively. We consider $\TmTw=1.1$, $1.2$ and $1.3$ for wall cooling, at wall temperatures of 277, 300 and 330\,K, and $\TmTw=0.9$, $0.85$ and $0.83$ for wall heating, at 330 and 360\,K. The resulting viscosity ratios are $\mu_m/\mu_w=0.21$--$0.64$ for wall cooling and $1.7$--$2.9$ for wall heating. The wall-to-mean viscosity ratio of the cooled cases, up to $4.8$, exceeds the range of the friction correction of \citet{petukhov1970} by more than a factor of two, whereas it lies within the range of his heat transfer correction (\S\,\ref{sec:fric}).

The resolution requirement is set by the temperature field. At $\Pran>1$ the smallest scale of the temperature is the Batchelor scale $\eta_B=\eta/\Pran^{1/2}$, with $\eta=(\bar\nu^3/\varepsilon)^{1/4}$ the Kolmogorov scale, and the meshes were sized so that $\Delta x/\eta_B\lesssim4.3$, $\Delta y/\eta_B\lesssim1.4$ and $\Delta z/\eta_B\lesssim2.9$. All cases were run until $C_f$ and $q_w$ were statistically stationary, after which the statistics were collected over at least $15$ eddy turnover times $h/u_\tau$. The residual of the integrated mean momentum balance (\S\,\ref{sec:balances}) is below $0.5\,\%$ of $\tau_w$ for $\ycp>30$ in all cases, where $\ycp=y^+\mu_w/\bar\mu$ is the semi-local wall distance and $\bar\mu$ the local mean viscosity.

%=====================================================================================================
\section{Instantaneous flow field}\label{sec:inst}

\begin{figure}
\centerline{\includegraphics[width=\textwidth]{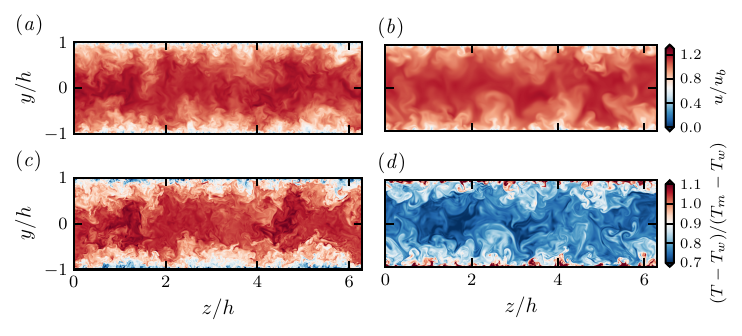}}
\caption{Instantaneous (\textit{a,b}) streamwise velocity, $u/u_b$, and (\textit{c,d}) temperature, $(T-T_w)/(T_m-T_w)$, in a cross-stream plane, for flow cases (\textit{a,c}) H12-B (wall cooling, $\TmTw=1.2$, $\Rey_\tau=501$, $\Retcp=1370$) and (\textit{b,d}) H083 (wall heating, $\TmTw=0.83$, $\Rey_\tau=891$, $\Retcp=317$). The colour scale of the temperature spans the core values and saturates in the wall layers. In panel (\textit{d}) the colour map is reversed with respect to the colourbar, so that red marks the hot fluid in both cases, at the core for wall cooling and at the wall for wall heating. \jfmnb{https://cocalc.ai/share/modesti-ceci-water-channel/Figure-01/figure01.ipynb}}
\label{fig:cross}
\end{figure}

Figure~\ref{fig:cross} shows the instantaneous velocity and temperature in a cross-stream plane for two flow cases, one with wall cooling (H12-B) and one with wall heating (H083). Both display the features of wall turbulence, with high-speed fluid from the core reaching the walls and low-speed fluid ejected towards the channel centre. However, the size of the eddies is very different. The cooled case has the lower friction Reynolds number at the wall ($\Rey_\tau=501$ against $891$) but the finer structures, with small eddies populating the whole section, whereas the heated case appears as a low-pass filtered version of it, dominated by large structures extending from the walls to the centre. Heating the wall therefore leads to a higher bulk flow viscosity than at the wall, which tends to lower the effective Reynolds number, suppressing finer structures. We note that the exact same trend is observed in air \citep{modesti2024air} when heating, despite the viscosity of air increasing with temperature. This can be attributed to the density variations in air, which scale as $T_w/T$ and therefore exceed the viscosity variations, which scale approximately as $(T/T_w)^{0.76}$, so that in air the density, rather than the viscosity, controls the response to heating and cooling, and it affects the outer layer as well. 
The temperature field (figure~\ref{fig:cross}\textit{c,d}) is well mixed in the core, and the largest temperature variations are confined to a thin conductive layer adjacent to the walls, as expected at $\Pran_w=2$--$6$. The temperature field also shows sharper eddy boundaries and finer details than the velocity. The absence of a pressure term in the energy equation makes the scalar interfaces sharper than the velocity ones even at $\Pran\le1$ \citep{pirozzoli2016}, and a Prandtl number larger than one adds finer scales.

\begin{figure}
\centerline{\includegraphics[width=\textwidth]{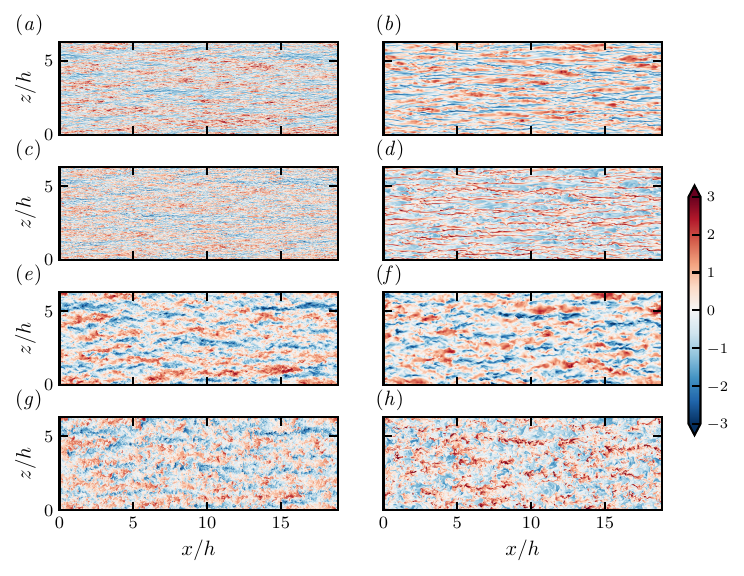}}
\caption{Instantaneous fluctuations of the streamwise velocity, $u'/u'_{rms}$, and of the temperature, $\theta'/\theta'_{rms}$, in wall-parallel planes, for flow cases (\textit{a,c,e,g}) H12-B (wall cooling) and (\textit{b,d,f,h}) H083 (wall heating). Panels (\textit{a--d}) are at the semi-local wall distance $\ycp=y^+\mu_w/\bar\mu=15$, corresponding to (\textit{a,c}) $y^+=7$ and (\textit{b,d}) $y^+=31$. Panels (\textit{e--h}) are at $y/h=0.2$, corresponding to (\textit{e,g}) $\ycp=255$, $y^+=100$ and (\textit{f,h}) $\ycp=72$, $y^+=179$. (\textit{a,b,e,f}) $u'/u'_{rms}$ and (\textit{c,d,g,h}) $\theta'/\theta'_{rms}$, with the colour scale spanning $-3$ to $3$. \jfmnb{https://cocalc.ai/share/modesti-ceci-water-channel/Figure-02/figure02.ipynb}}
\label{fig:wallpar}
\end{figure}

Wall-parallel planes are shown in figure~\ref{fig:wallpar}. Since the wall viscosity is not representative of the near-wall region, we place the near-wall plane at the semi-local wall distance $\ycp=y^+\mu_w/\bar\mu=15$. This distance corresponds to $y^+=7$ for wall cooling and to $y^+=31$ for wall heating. At this distance both flows exhibit the alternating high- and low-speed streaks of the buffer layer (figure~\ref{fig:wallpar}\textit{a,b}). In outer units the streaks of the cooled case are much finer and are superposed on patches of size $h$, whereas in the heated case they are coarser and more coherent. The temperature fluctuations (figure~\ref{fig:wallpar}\textit{c,d}) follow the velocity streaks closely, and the correlation coefficient between $u'$ and $\theta'$ in the plane is $+0.85$ for wall cooling and $-0.85$ for wall heating. The sign reflects the direction of the heat flux, as low-speed fluid lifted from the wall carries the wall temperature into the layer. For wall heating the temperature is strongly skewed, with narrow hot filaments located on the low-speed streaks. Further from the wall, at $y/h\approx0.2$ (figure~\ref{fig:wallpar}\textit{e--h}), the velocity is organised in streamwise-elongated regions of width comparable to $h$ in both cases, while the temperature is more fragmented at small scales and its correlation with the velocity drops to about $\pm0.6$.

%=====================================================================================================
\section{Spectral analysis}\label{sec:spectra}

\begin{figure}
\centerline{\includegraphics[width=\textwidth]{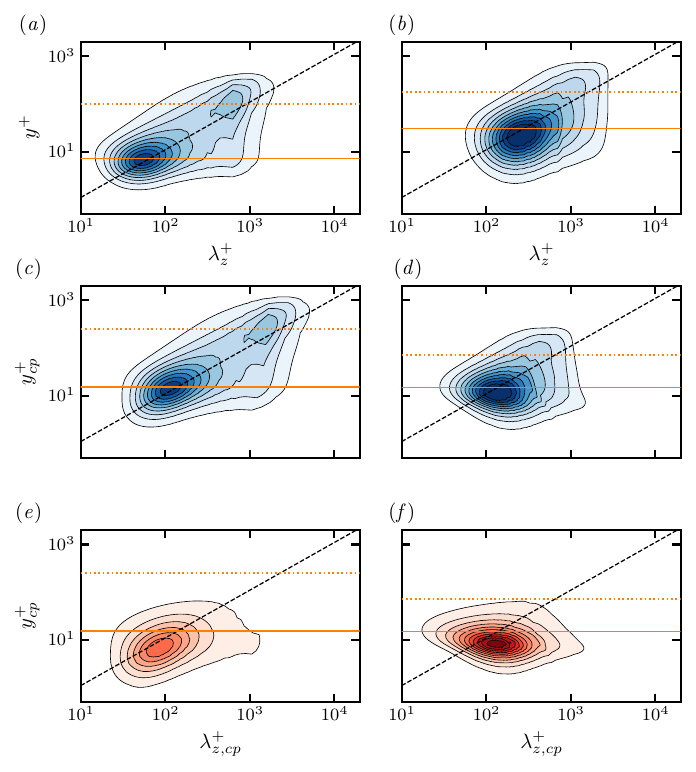}}
\caption{Premultiplied spanwise spectra of (\textit{a--d}) the streamwise velocity, $k_zE_u/u_\tau^2$, and (\textit{e,f}) the temperature, $k_zE_\theta/\theta_\tau^2$, for flow cases H12-B (left) and H083 (right). Panels (\textit{a,b}) are in wall units, $\lambda_z^+$ and $y^+$, and panels (\textit{c--f}) in semi-local units, $\lamcp=\lambda_zu_\tau/\bar\nu$ and $\ycp$. The velocity spectrogram contour levels go from $0.36$ to $3.6$ in steps of $0.36$, and the temperature spectrogram levels are twelve equal intervals of the common maximum. The orange lines mark the positions of the wall-parallel planes of figure~\ref{fig:wallpar} (solid, $\ycp\approx15$, and dotted, $y/h\approx0.2$), and the dashed line is $y=0.11\lambda_z$. \jfmnb{https://cocalc.ai/share/modesti-ceci-water-channel/Figure-03/figure03.ipynb}}
\label{fig:spectrograms}
\end{figure}

The visual impression of figures~\ref{fig:cross} and \ref{fig:wallpar} is made quantitative by the premultiplied spanwise spectra $k_zE_u$ and $k_zE_\theta$, where $E_u(k_z,y)$ and $E_\theta(k_z,y)$ are the spectral densities of the streamwise velocity and temperature fluctuations, $k_z$ is the spanwise wavenumber and $\lambda_z=2\upi/k_z$ the corresponding wavelength. The spectra are shown in figure~\ref{fig:spectrograms} for the same two flow cases, both in wall units and in semi-local units. In the latter, lengths are scaled with the local viscous length $\bar\nu/u_\tau$, so that $\lamcp=\lambda_zu_\tau/\bar\nu$ and $\ycp=yu_\tau/\bar\nu$, which coincides with the constant-property coordinate of \S\,\ref{sec:transf}. When scaled in wall units (figure~\ref{fig:spectrograms}\textit{a,b}) the spectrograms of the streamwise velocity are very different. For wall cooling the inner peak sits at $y^+\approx7$ and $\lambda_z^+\approx55$, whereas for wall heating it sits at $y^+\approx22$ and $\lambda_z^+\approx230$, a factor of four larger than for wall cooling and twice the $\lambda_z^+\approx100$ of constant-property flow. These are the positions and spacings of the streaks seen in the wall-parallel planes at $\ycp=15$ (solid orange lines). In semi-local units (figure~\ref{fig:spectrograms}\textit{c,d}), with the local viscosity $\bar\nu(y)$ in the definition of both the wavelength and the wall distance, the inner peak of both cases moves to $\ycp\approx11$--$14$ and $\lamcp\approx120$, as in constant-property flow. Across the whole database the inner site lies at $\ycp=11$--$15$ and $\lamcp=110$--$140$, for wall heating and wall cooling alike. The streaks of figure~\ref{fig:wallpar}(\textit{a,b}) therefore have the same spacing in local viscous units, and they look different only because at $\ycp=15$ the local viscous length differs by a factor of four between the two cases. The scale separation is also different. For wall cooling ($\Retcp=\ycp(h)=1370$) the spectrogram extends along the ridge $y\approx0.11\lambda_z$ of the attached eddies up to an outer site at $\lambda_z\approx1.3h$ and $y/h\approx0.15$--$0.4$, which corresponds to the large patches visible at $y/h=0.2$. For wall heating ($\Retcp=317$) no distinct outer site is present and the spectrum is confined to $\lamcp\lesssim10^3$, which is why the large scales dominate the cross-stream view of figure~\ref{fig:cross}(\textit{b}).

The temperature spectra (figure~\ref{fig:spectrograms}\textit{e,f}) peak at a spanwise wavelength comparable to that of the velocity, $\lamcp\approx75$--$95$ for wall cooling and $115$--$145$ for wall heating, but closer to the wall, at $\ycp\approx5$--$8$ and $\ycp\approx7$--$9$, respectively. At $\Pran_w=2$--$12$ the temperature fluctuations are indeed concentrated in the conductive sublayer and in the buffer layer, where the mean temperature gradient is largest. In both cases the energy decays rapidly away from the wall and no outer site appears, apart from a weak tail towards $\lamcp\sim10^3$ for wall cooling. At $y/h=0.2$ the temperature therefore carries less large-scale energy than the velocity, consistent with the fragmented appearance of figure~\ref{fig:wallpar}(\textit{g,h}).

\begin{figure}
\centerline{\includegraphics[width=\textwidth]{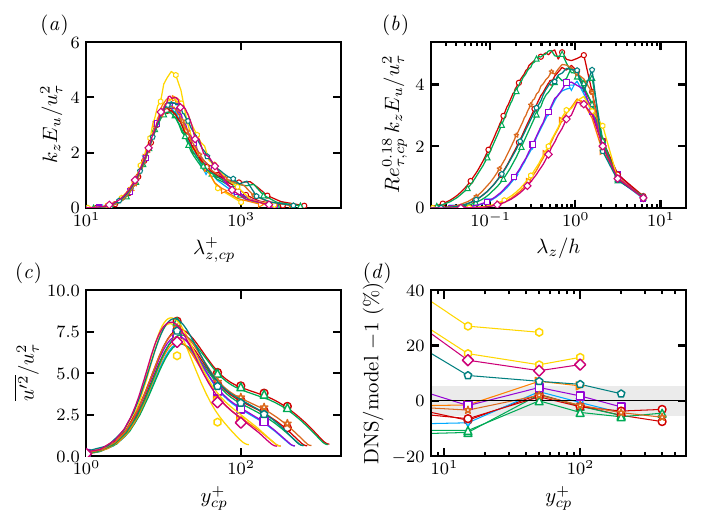}}
\caption{(\textit{a}) Premultiplied spanwise spectra of the streamwise velocity at $\ycp=15$ against $\lamcp$. (\textit{b}) Spectra at $\ycp=100$, compensated by $\Retcp^{0.18}$, against $\lambda_z/h$, for the cases with $\Retcp>170$. (\textit{c}) Streamwise velocity variance against $\ycp$ (lines), with the model of \citet{pirozzoli2024variance} evaluated at $\Retcp$ (symbols). (\textit{d}) Relative deviation of the DNS from the model, the grey band denoting $\pm5\,\%$. Colours and symbols denote the property conditions, namely \condkey. \jfmnb{https://cocalc.ai/share/modesti-ceci-water-channel/Figure-04/figure04.ipynb}}
\label{fig:specvar}
\end{figure}

The semi-local scaling of the inner site holds across the whole database (figure~\ref{fig:specvar}\textit{a}). At $\ycp=15$ the spectra of all cases nearly coincide for $\lamcp\lesssim300$, while at larger wavelengths they separate according to $\Retcp$, which controls the extent of the outer range. \citet{pirozzoli2024variance} argued that in constant-property flows the large-scale part of the spectrum scales in outer units with amplitude proportional to $\Rey_\tau^{-0.18}$, so that the streamwise variance at a fixed inner-scaled distance follows $\overline{u'^2}^+=A-B\Rey_\tau^{-0.18}$. Using $\Retcp$ in place of $\Rey_\tau$, the outer-scaled spectra at $\ycp=100$ (figure~\ref{fig:specvar}\textit{b}) collapse for $\lambda_z\gtrsim h$, for heated and cooled walls alike, whereas the intermediate scales do not, as no overlap range can be identified at these Reynolds numbers. The model of \citet{pirozzoli2024variance} with $\ycp$ and $\Retcp$ predicts the variance of the cooled cases within $-8\,\%$ to $+7\,\%$ for $\ycp=50$--$400$ (figure~\ref{fig:specvar}\textit{c,d}), with a deficit at the buffer-layer peak that grows with the property contrast, from $-2$ to $-3\,\%$ at $\TmTw=1.1$ to $-11\,\%$ at $\TmTw=1.3$. For wall heating the variance is above the model, by $+9\,\%$ at $\TmTw=0.9$, by $+15$--$17\,\%$ at $\TmTw=0.83$--$0.85$ and by $+27\,\%$ for L083, the case at the lowest $\Retcp$. The excess persists for H09 ($\Retcp=630$), which is inside the calibration range of the model, so that it is not only a low-Reynolds-number effect. Hence semi-local scaling accounts for the size of the near-wall eddies, but not entirely for their intensity, in line with the findings of \citet{patel2016} for variable-viscosity flows.

%=====================================================================================================
\section{Mean flow field and variable-property transformations}\label{sec:mean}

\subsection{Mean profiles in wall units and in defect form}

\begin{figure}
\centerline{\includegraphics[width=\textwidth]{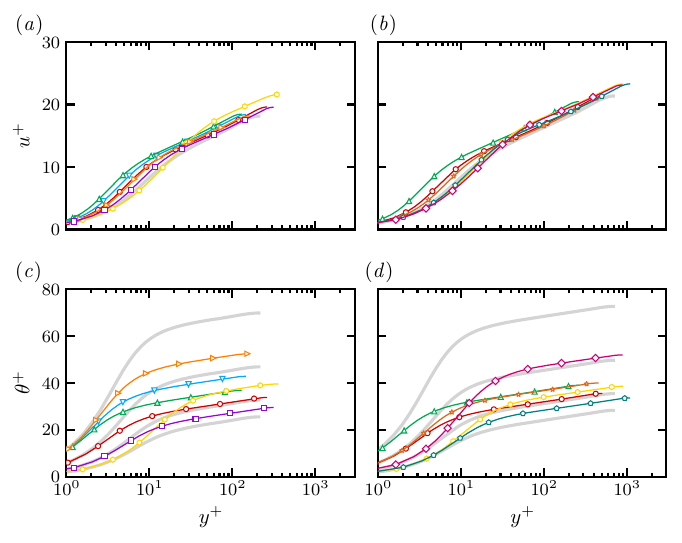}}
\caption{Mean (\textit{a,b}) velocity and (\textit{c,d}) temperature profiles in wall units, for (\textit{a,c}) L and (\textit{b,d}) H flow cases. The grey lines are the constant-property profiles of \citet{pirozzoli2024profiles} at the median $\Rey_\tau$ of each group and, for the temperature, at each wall Prandtl number. Colours and symbols as in figure~\ref{fig:specvar}. \jfmnb{https://cocalc.ai/share/modesti-ceci-water-channel/Figure-05/figure05.ipynb}}
\label{fig:raw}
\end{figure}

The mean velocity and temperature profiles in wall units (figure~\ref{fig:raw}) are compared with the constant-property composite profiles of \citet{modesti2024air} for the velocity and of \citet{pirozzoli2024profiles} for the temperature, which are obtained by matching the inner-layer distributions derived from the eddy viscosity of \citet{musker1979} and the eddy diffusivity of \citet{pirozzoli2023} with parabolic outer distributions. At constant properties these profiles reproduce the DNS of passive scalars in channels \citep{pirozzoli2016,modesti2024air}. As in air, the variable-property profiles depart from the reference starting from the buffer layer, with changes of intercept of the logarithmic layer, so that no universal law of the wall exists in wall units. For wall cooling the velocity lies above the reference in the buffer layer and below it in the logarithmic layer, and for wall heating the opposite occurs. The temperature in wall units depends on the wall Prandtl number, which sets the thickness of the conductive sublayer and the intercept of the thermal logarithmic law, and on the variation of $\Pran$ across the layer, which has no counterpart in constant-property flow.

\begin{figure}
\centerline{\includegraphics[width=\textwidth]{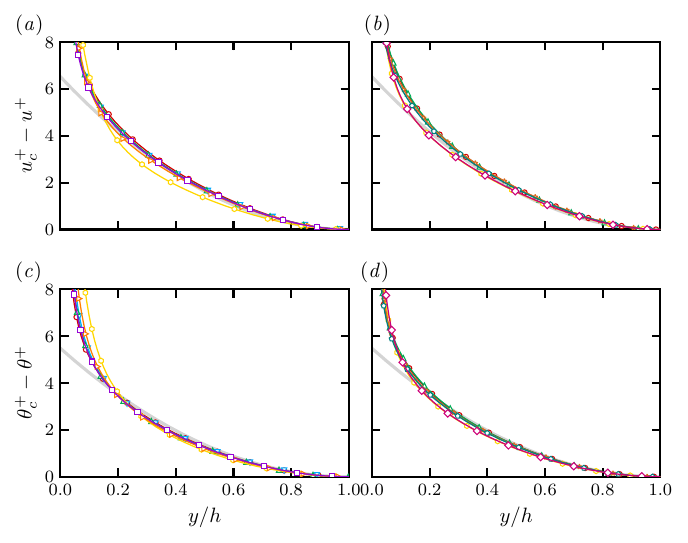}}
\caption{Mean (\textit{a,b}) velocity and (\textit{c,d}) temperature defect profiles, for (\textit{a,c}) L and (\textit{b,d}) H flow cases. The grey lines are the parabolic core distributions of the constant-property model, $u_c^+-u^+=6.516(1-y/h)^2$ for the velocity, with the constant determined by fitting the channel DNS of \citet{leemoser2015}, and $\theta_c^+-\theta^+=5.48(1-y/h)^2$ for the temperature \citep{pirozzoli2024profiles}. Colours and symbols as in figure~\ref{fig:specvar}. \jfmnb{https://cocalc.ai/share/modesti-ceci-water-channel/Figure-06/figure06.ipynb}}
\label{fig:defect}
\end{figure}

A remarkable simplification occurs in the outer layer. Since the density is constant and the heat source is uniform, the integrated mean momentum and thermal balances (\S\,\ref{sec:balances}) reduce, away from the wall where the molecular terms are negligible, to $-\overline{u'v'}\simeq u_\tau^2(1-\eta)$ and $-\overline{v'\theta'}\simeq u_\tau\theta_\tau(1-\eta)$, with $\eta=y/h$. In the core the eddy viscosity scales with $u_\tau h$ and not with the local molecular properties, as in the uniform outer eddy viscosity of \citet{clauser1956}, and the same holds for the eddy diffusivity \citep{pirozzoli2024profiles}, so that the velocity and temperature defects in wall units should be independent of the property variations. In air this is not the case, because the density variation enters both the total heat flux and the definition of the friction velocity \citep{modesti2024air}, so that a density-weighted defect would be needed. Figure~\ref{fig:defect} confirms the outer-layer similarity of the defects $u_c^+-u^+$ and $\theta_c^+-\theta^+$, where the subscript $c$ denotes the centreline value. The profiles of the H cases follow each other and the constant-property parabolas closely for $y/h\gtrsim0.15$. At $y/h=0.2$--$0.4$ the velocity defect is within $-4$ to $+9\,\%$ of the parabola for wall cooling, and the temperature defect is within $-13$ to $+5\,\%$ for all cases but L083 ($-16\,\%$). The largest departure, $-10$ to $-19\,\%$ in the velocity defect, is found for L083, the heated case at the lowest equivalent Reynolds number ($\Retcp=125$), where the outer layer is not well separated from the wall layer. Outer-layer similarity of the untransformed defects implies that the mean property variation affects the bulk velocity and the mixed mean temperature only through the wall layer.

\subsection{Mean momentum and thermal balances}\label{sec:balances}

\begin{figure}
\centerline{\includegraphics[width=\textwidth]{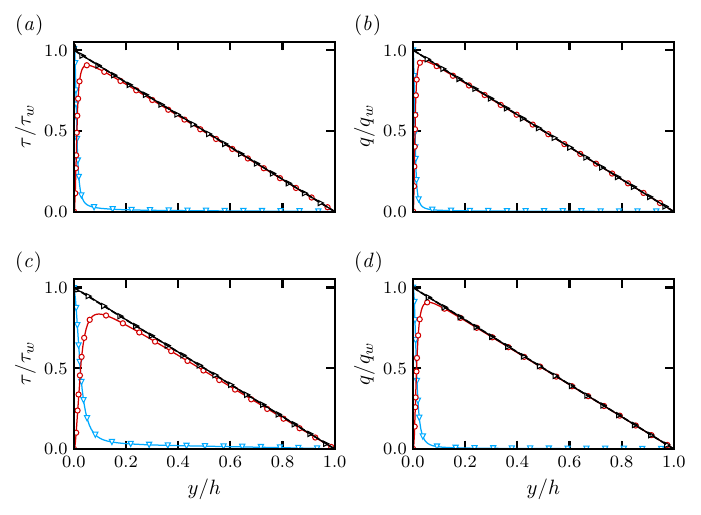}}
\caption{Integrated mean (\textit{a,c}) momentum and (\textit{b,d}) thermal balances \eqref{eq:mmb}--\eqref{eq:mtb}, for flow cases (\textit{a,b}) H12-B (wall cooling) and (\textit{c,d}) H083 (wall heating). The symbols indicate viscous stress or mean conduction (blue downward triangles), turbulent stress or heat flux (red circles), and the sum of the two (black right triangles). The dashed line is the total, $1-y/h$. \jfmnb{https://cocalc.ai/share/modesti-ceci-water-channel/Figure-07/figure07.ipynb}}
\label{fig:balances}
\end{figure}

Integrating the mean momentum and energy equations once in $y$ yields
\begin{align}
\bar\mu\,\frac{\mathrm{d}\bar u}{\mathrm{d}y}+\overline{\mu'\frac{\partial u'}{\partial y}}-\rho\,\overline{u'v'}&=\tau_w\,(1-\eta),\label{eq:mmb}\\
\bar\lambda\,\frac{\mathrm{d}\bar T}{\mathrm{d}y}+\overline{\lambda'\frac{\partial T'}{\partial y}}-\rho c_p\,\overline{v'T'}&=q_w\,(1-\eta).\label{eq:mtb}
\end{align}
The terms of \eqref{eq:mmb}--\eqref{eq:mtb} are shown in figure~\ref{fig:balances} for one case with wall cooling and one with wall heating. The balances are qualitatively similar to those of a passive scalar, with the molecular terms dominating near the wall and the turbulent fluxes elsewhere. Away from the wall the sum of the sampled molecular and turbulent terms matches the total within $0.5\,\%$, which also certifies the statistical convergence. The fluctuating-property terms are not sampled during the simulations and they are not shown. For wall heating it is positive and further from the wall, $+1.3$ to $+2.4\,\%$ at $y^+\approx5$--$6$. The sign follows from $\mathrm{d}\mu/\mathrm{d}T<0$ and from the correlation between $u'$ and $T'$ discussed in \S\,\ref{sec:inst}. The thermal term is an order of magnitude smaller, $|\overline{\lambda'\partial T'/\partial y}|\le0.8\,\%$ of $q_w$, because the conductivity varies much less than the viscosity. Both terms are neglected in the derivation of the transformations.

\subsection{Variable-property transformations for liquids}\label{sec:transf}

Following \citet{modesti2024air}, we seek integral transformations
\begin{equation}
\ycp=\int_0^{y^+}f\,\mathrm{d}y^+,\qquad u_{cp}^+=\int_0^{u^+}g\,\mathrm{d}u^+,\qquad \theta_{cp}^+=\int_0^{\theta^+}h\,\mathrm{d}\theta^+,\label{eq:transf}
\end{equation}
that map the variable-property profiles onto those of a constant-property (cp) flow with passive scalar at the wall Prandtl number $\Pran_w$. Let $N=\bar\mu/\mu_w=\bar\nu/\nu_w$, $L=\bar\lambda/\lambda_w$ and $\Pran=\Pran_wN/L$ the local Prandtl number. Introducing an eddy viscosity and an eddy diffusivity, $-\overline{u'v'}=\nu_T\,\mathrm{d}\bar u/\mathrm{d}y$ and $-\overline{v'T'}=\alpha_T\,\mathrm{d}\bar T/\mathrm{d}y$, and neglecting the fluctuating-property terms, equations \eqref{eq:mmb}--\eqref{eq:mtb} become in wall units
\begin{equation}
N\left(1+\frac{\nu_T}{\bar\nu}\right)\frac{\mathrm{d}u^+}{\mathrm{d}y^+}=1-\eta,\qquad
\frac{L}{\Pran_w}\left(1+\Pran\,\frac{\alpha_T}{\bar\nu}\right)\frac{\mathrm{d}\theta^+}{\mathrm{d}y^+}=1-\eta,
\label{eq:balances}
\end{equation}
to be compared with their constant-property counterparts, $(1+\nu_{T,cp}/\nu)\,\mathrm{d}u_{cp}^+/\mathrm{d}\ycp=1-\eta$ and $\Pran_w^{-1}(1+\Pran_w\,\alpha_{T,cp}/\nu)\,\mathrm{d}\theta_{cp}^+/\mathrm{d}\ycp=1-\eta$. 
Enforcing the semi-local stretching used in \S\S\,\ref{sec:inst}--\ref{sec:spectra} ($\ycp=y^+/N=yu_\tau/\bar\nu$) fixes the kernel for the wall distance $f=\mathrm{d}(y/N)/\mathrm{d}y$
and substituting the transformations~\eqref{eq:transf} into~\eqref{eq:balances} we find the remaining kernels,
\begin{equation}
f=\frac{\mathrm{d}}{\mathrm{d}y}\left(\frac{y}{N}\right),\qquad
g=Nf\,\frac{1+\nu_T/\bar\nu}{1+\nu_{T,cp}/\nu},\qquad
h=Lf\,\frac{1+\Pran\,\alpha_T/\bar\nu}{1+\Pran_w\,\alpha_{T,cp}/\nu}.\label{eq:kernels}
\end{equation}
The velocity kernel is the constant-density limit of the one derived for air, and it is formally equivalent to the transformation of \citet{hasan2023}. The temperature kernel differs from the air version, accounting for the variable Prandtl number. In the viscous and conductive sublayers the kernels reduce to $g\simeq Nf$ and $h\simeq Lf$, whereas in the turbulent region both tend to $Nf\simeq1$, consistently with the outer-layer similarity of the defects (figure~\ref{fig:defect}). The transition between the two limits occurs at $\ycp\approx2$--$10$ for $\Pran_w=2$--$12$, and it is the eddy-diffusivity weighting in \eqref{eq:kernels} that blends them. 

The equivalent channel height $h_{cp}=\int_0^hf\,\mathrm{d}y=h\,\mu_w/\bar\mu_c$ defines the equivalent friction Reynolds number
\begin{equation}
\Retcp=\frac{h_{cp}}{\delta_v}=\Rey_\tau\,\frac{\mu_w}{\bar\mu_c},\label{eq:retcp}
\end{equation}
with $\bar\mu_c$ the mean viscosity at the centreline. $\Retcp$ is larger than $\Rey_\tau$ for wall cooling and smaller for wall heating, and it is the parameter that controls the scale separation of \S\S\,\ref{sec:inst}--\ref{sec:spectra}.

\begin{figure}
\centerline{\includegraphics[width=\textwidth]{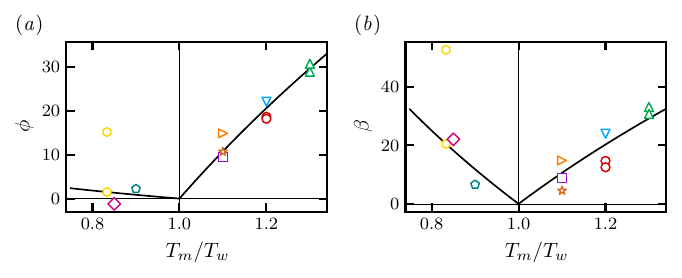}}
\caption{Corrections (\textit{a}) $\phi$ and (\textit{b}) $\beta$ of the eddy viscosity and eddy diffusivity \eqref{eq:eddy} that reproduce the DNS friction and heat transfer coefficients, as a function of $\TmTw$. The solid lines are the fit \eqref{eq:law}. Colours and symbols as in figure~\ref{fig:specvar}. \jfmnb{https://cocalc.ai/share/modesti-ceci-water-channel/Figure-08/figure08.ipynb}}
\label{fig:coeffs}
\end{figure}

As in \citet{modesti2024air}, the eddy viscosity and diffusivity are modelled as functions of $\ycp$,
\begin{equation}
\frac{\nu_T}{\bar\nu}=\frac{(\kappa\ycp)^3}{(\kappa\ycp)^2+C_{v1}^2+\phi},\qquad
\frac{\alpha_T}{\bar\nu}=\frac{(\kappa_\theta\ycp)^3}{(\kappa_\theta\ycp)^2+C_{\theta}^2+\beta},\label{eq:eddy}
\end{equation}
with $\kappa=0.387$ \citep{pirozzoli2021}, $\kappa_\theta=0.459$ and $C_\theta=10$ \citep{pirozzoli2023} and $C_{v1}=7.1$, determined by fitting the channel DNS of \citet{leemoser2015}. With $\phi=\beta=0$ the eddy viscosity reduces to the constant-property model of \citet{musker1979}, and the eddy diffusivity to its counterpart for the temperature \citep{pirozzoli2023}, which are the cp reference. The corrections $\phi$ and $\beta$ account empirically for the property variations, and are determined from the DNS data. The values obtained are shown in figure~\ref{fig:coeffs}. For wall cooling both corrections grow with $\TmTw$. For wall heating $\phi$ is close to zero, so that the eddy viscosity requires essentially no correction, while $\beta$ is positive. We fit the data with one coefficient per branch, vanishing at $\TmTw=1$,
\begin{equation}
\phi=\begin{cases}113\,\ln(\TmTw), & \TmTw>1,\\ -8.2\,\ln(\TmTw), & \TmTw<1,\end{cases}\qquad
\beta=\begin{cases}111\,\ln(\TmTw), & \TmTw>1,\\ -112\,\ln(\TmTw), & \TmTw<1,\end{cases}\label{eq:law}.
\end{equation}
A small improvement on the prediction of $\phi$ and $\beta$, can be achieved using the mean-to-wall viscosity ratio $\phi\simeq-19.0\ln(\mu_m/\mu_w)$ and $\beta\simeq-18.5\ln(\mu_m/\mu_w)$, which reduce the root-mean-square scatter of the per-case values from $2.0$ to $0.9$ for $\phi$ and from $4.7$ to $3.6$ for $\beta$. We nevertheless retain \eqref{eq:law}, which replicates the laws for air \citep{modesti2024air}.

\begin{figure}
\centerline{\includegraphics[width=\textwidth]{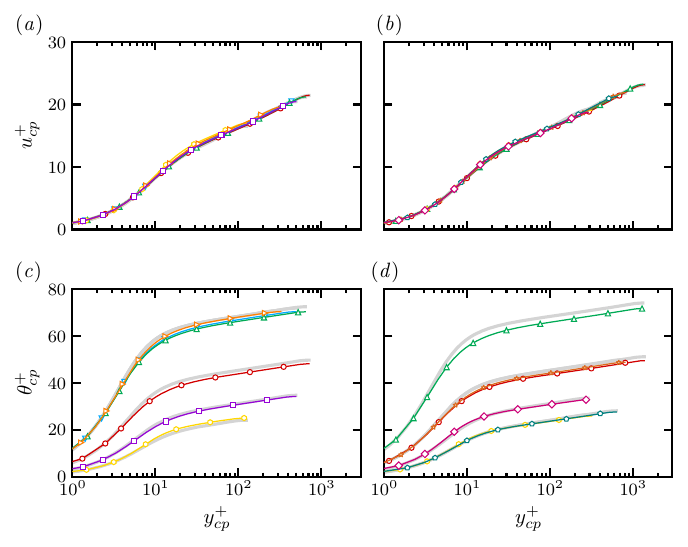}}
\caption{Mean (\textit{a,b}) velocity and (\textit{c,d}) temperature profiles transformed with \eqref{eq:transf}--\eqref{eq:law}, for (\textit{a,c}) L and (\textit{b,d}) H flow cases. The grey lines are the constant-property profiles \citep{pirozzoli2024profiles} at the $\Retcp$ and $\Pran_w$ of each case. Colours and symbols as in figure~\ref{fig:specvar}. \jfmnb{https://cocalc.ai/share/modesti-ceci-water-channel/Figure-09/figure09.ipynb}}
\label{fig:transformed}
\end{figure}

The transformed profiles are shown in figure~\ref{fig:transformed}. The velocity profiles closely follow the constant-property reference for all cases, and so do the temperature profiles, which group into four families according to the wall Prandtl number. Small residual deviations remain in the temperature for wall cooling, where $\theta_{cp}^+$ lies $1$--$2$ units below the reference in the logarithmic region. They are due to the near-wall eddy diffusivity, which in the DNS is about twice the model \eqref{eq:eddy} at $\ycp\lesssim5$ and slightly smaller than it at $\ycp\approx20$--$100$, whereas a constant $\beta$ shifts the model in one direction only and cannot match both regions. 

%=====================================================================================================
\section{Wall friction and heat transfer}\label{sec:fric}

The transformations \eqref{eq:transf}--\eqref{eq:law} can be inverted to predict the mean velocity and temperature profiles, and hence the friction and heat transfer coefficients, for given $\Rey_b$, $\TmTw$ and $T_w$. We use the iterative procedure of \citet{modesti2024air} (their algorithm~1). Starting from the constant-property profiles of \citet{pirozzoli2024profiles} at an estimate of $\Retcp$ and at $\Pran_w$, the inverse integrals $y=\int f^{-1}\mathrm{d}\ycp$, $\bar u=\int g^{-1}\mathrm{d}u_{cp}$ and $\theta=\int h^{-1}\mathrm{d}\theta_{cp}$ are evaluated, the kernels are updated with the resulting temperature through the property laws \eqref{eq:mu}--\eqref{eq:lam}, and the procedure is repeated, together with the update of $\Retcp$, until the target bulk Reynolds number and temperature ratio are matched. The properties are evaluated at the mean temperature, $\mu(\bar T)$ and $\lambda(\bar T)$, which is all the inverse procedure can access. In the DNS the sampled $\overline{\mu(T)}$ exceeds $\mu(\bar T)$ by at most $3.5\,\%$, in the buffer layer of the cases with the largest property contrast.

\begin{figure}
\centerline{\includegraphics[width=\textwidth]{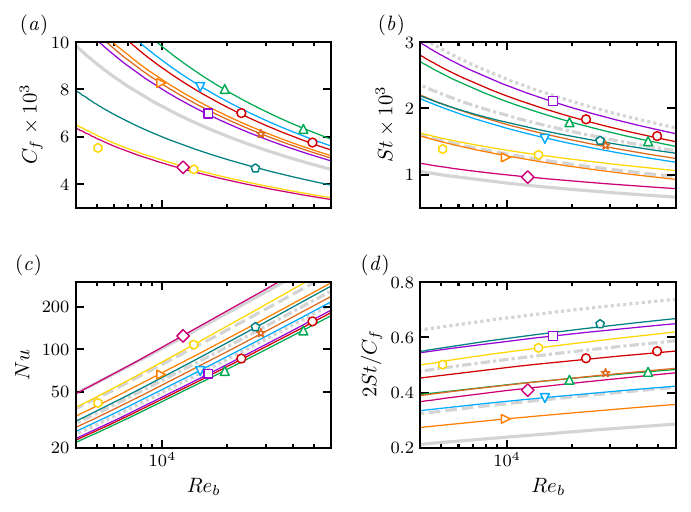}}
\caption{(\textit{a}) Friction coefficient, (\textit{b}) Stanton number, (\textit{c}) Nusselt number and (\textit{d}) Reynolds analogy factor as functions of the bulk Reynolds number. The symbols are the DNS data and the coloured lines the predictions obtained by inverting the transformations \eqref{eq:transf}--\eqref{eq:law} at the property conditions of the DNS. The grey lines are the constant-property values obtained from the composite profiles of \S\,\ref{sec:mean}, at $\Pran=11.58$ (solid), $5.93$ (dashed), $3.16$ (dash-dotted) and $2.02$ (dotted) for the thermal quantities. Colours and symbols as in figure~\ref{fig:specvar}. \jfmnb{https://cocalc.ai/share/modesti-ceci-water-channel/Figure-10/figure10.ipynb}}
\label{fig:coef}
\end{figure}

The DNS coefficients are compared with the predictions in figure~\ref{fig:coef}. With respect to constant-property flow at the same $\Rey_b$ (based on $\mu_m$), wall cooling increases the friction coefficient, by up to $28$--$32\,\%$ at $\TmTw=1.3$, while wall heating reduces it, by up to $30$--$39\,\%$ at $\TmTw=0.83$--$0.85$, approximately matching the $15$--$26\,\%$ drag reduction reported by \citet{lee2013} in boundary layers at $T_\infty/T_w=0.81$--$0.88$. The sign of the effect is the same as in air \citep{modesti2024air}, where wall heating also reduces the friction at a given $\Rey_b$, although there the viscosity varies in the opposite direction and the density is not constant. The Stanton number depends strongly on the reference Prandtl number. With respect to the constant-property value at $\Pran_w$, $St$ is larger by a factor of $1.2$--$2.3$ for wall cooling and smaller by a factor of $0.48$--$0.76$ for wall heating. The Nusselt number compresses these differences in logarithmic scale and the Stanton number should be preferred for an accurate evaluation of theories. The Reynolds analogy factor $2St/C_f$ (figure~\ref{fig:coef}\textit{d}) is far from its constant-property value at the wall Prandtl number, since the Stanton number responds to the property variations much more strongly than the friction coefficient, so that $2St/C_f$ is larger by $12$--$73\,\%$ for wall cooling and smaller by $9$--$23\,\%$ for wall heating, compared to the passive scalar case. This departure from the Reynolds analogy is of interest in itself. More heat transfer for less friction is a goal of turbulence control, which is hard to reach because the same eddies transport heat and momentum, and it has been pursued by shaping the velocity field, for which \citet{motoki2018} determined the optimal states between parallel plates. Wall cooling of a liquid obtains it from the property variations alone, since the Stanton number rises more than the friction coefficient.

\begin{figure}
\centerline{\includegraphics[width=\textwidth]{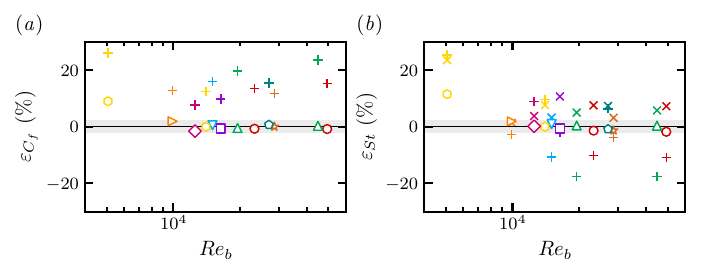}}
\caption{Percentage error of the predicted (\textit{a}) friction coefficient and (\textit{b}) Stanton number with respect to the DNS, as a function of $\Rey_b$, for the present method (open symbols), \citet{petukhov1970} ($+$) and \citet{siedertate1936} ($\times$). The grey band denotes $\pm2\,\%$. Colours as in figure~\ref{fig:specvar}. \jfmnb{https://cocalc.ai/share/modesti-ceci-water-channel/Figure-11/figure11.ipynb}}
\label{fig:errors}
\end{figure}

The predictions of the inverse transformations (coloured lines in figure~\ref{fig:coef}) follow the DNS for all property conditions. The prediction errors are reported in figure~\ref{fig:errors} and table~\ref{tab:errors}, together with those of the correlations of \citet{petukhov1970} and \citet{siedertate1936}, applied as pipe correlations with the hydraulic diameter $2h$ and the properties at $T_m$. For all eleven cases with $\Retcp\ge289$ the present method predicts $C_f$ and $St$ within $\pm2\,\%$, at four wall temperatures and for $\TmTw=0.85$--$1.3$, with no trend with $\Rey_b$. Petukhov's formula over-predicts the friction by $10$--$24\,\%$ for wall cooling and $8$--$16\,\%$ for wall heating, and under-predicts the heat transfer by up to $18\,\%$ for wall cooling. The correction of Sieder \& Tate over-predicts $St$ by $1$--$11\,\%$ for these cases, and both correlations over-predict $St$ by $24$--$25\,\%$ for the heated case at the lowest Reynolds number, L083. The errors of the heat transfer corrections are not a matter of extrapolation, since both were fitted to data that cover the present viscosity ratios and Prandtl numbers, apart from the lowest Reynolds numbers. The friction correction of \citet{petukhov1970}, instead, rests on water data with a wall-to-mean viscosity ratio of at most two, which most of the cooled cases exceed, and for these its error grows with the viscosity contrast, whereas the heated cases lie within its range and the error is still sizeable.

\begin{table}
\begin{center}
\def~{\hphantom{0}}
\begin{tabular}{lrrrrrrr}
\toprule
 & & & \multicolumn{2}{c}{$\varepsilon_{C_f}$ (\%)} & \multicolumn{3}{c}{$\varepsilon_{St}$ (\%)}\\
\cmidrule(lr){4-5}\cmidrule(lr){6-8}
Case & $\phi$ & $\beta$ & present & Petukhov & present & Petukhov & Sieder--Tate\\
\midrule
L12-B & 20.6 & 20.2 & -0.7 & +13.6 & -1.4 & -10.1 & +7.6 \\
L12-A & 20.6 & 20.2 & +0.5 & +16.0 & +1.0 & -10.7 & +3.2 \\
L13 & 29.6 & 29.1 & -0.4 & +19.7 & +0.5 & -17.6 & +5.0 \\
L083 & 1.5 & 20.4 & +9.0 & +26.1 & +11.5 & +25.4 & +23.6 \\
L11-A & 10.8 & 10.6 & +2.0 & +12.8 & +1.9 & -2.6 & +1.4 \\
L11-C & 10.8 & 10.6 & -0.6 & +9.8 & -0.6 & -1.8 & +10.7 \\
H12-B & 20.6 & 20.2 & -0.8 & +15.3 & -1.8 & -10.8 & +7.3 \\
H13 & 29.6 & 29.1 & +0.3 & +23.7 & +0.4 & -17.7 & +5.8 \\
H083 & 1.5 & 20.4 & +0.0 & +12.6 & +0.0 & +9.7 & +7.7 \\
H11-B & 10.8 & 10.6 & +0.1 & +11.8 & -1.3 & -3.8 & +3.2 \\
H09 & 0.9 & 11.8 & +0.7 & +15.5 & -0.8 & +6.4 & +7.4 \\
H085 & 1.3 & 18.2 & -1.5 & +7.7 & +0.2 & +8.9 & +3.8 \\
\bottomrule
\end{tabular}
\caption{Corrections \eqref{eq:law} and percentage errors of the predicted friction coefficient and Stanton number with respect to the DNS.}
\label{tab:errors}
\end{center}
\end{table}

The only case outside the $\pm2\,\%$ band is the heated L case L083, for which $C_f$ is over-predicted by $9\,\%$ and $St$ by $12\,\%$. This flow has $\Retcp=125$, so that its wall layer is effectively that of a constant-property channel where no logarithmic layer exists. Consistently, the transformed bulk velocity of L083 differs from the constant-property model by the same amount. A low-Reynolds-number correction of the reference profiles could remove this outlier. The engineering correlations are also outside their range there ($\Rey_b<10^4$).

%=====================================================================================================
\section{Application to the Spalart--Allmaras model}\label{sec:sa}

The eddy-viscosity and eddy-diffusivity corrections of \S\,\ref{sec:transf} can be carried over to RANS models. We consider the one-equation model of \citet{spalart1994} with the heat-transfer extension proposed by \citet{ceci2025}, which improves the near-wall asymptotic behaviour of both the eddy viscosity and the eddy diffusivity, and we solve the steady, fully developed channel flow with the property laws \eqref{eq:mu}--\eqref{eq:lam}. The working variable $\tilde\nu$ obeys
\begin{equation}
0=c_{b1}\tilde S\tilde\nu-c_{w1}f_w\left(\frac{\tilde\nu}{y}\right)^2+\frac{1}{\sigma}\left\{\frac{\mathrm{d}}{\mathrm{d}y}\left[(\bar\nu+\tilde\nu)\frac{\mathrm{d}\tilde\nu}{\mathrm{d}y}\right]+c_{b2}\left(\frac{\mathrm{d}\tilde\nu}{\mathrm{d}y}\right)^2\right\},\label{eq:sa}
\end{equation}
with $\bar\nu=\nu(\bar T)$ the local kinematic viscosity at the mean temperature, $\chi=\tilde\nu/\bar\nu$, and the standard functions and constants of the model \citep{spalart1994}, except $\kappa=0.387$. In the standard model the eddy viscosity is $\nu_t=\tilde\nu f_{v1}$, with $f_{v1}=\chi^3/(\chi^3+c_{v1}^3)$ and $c_{v1}=7.1$, and the eddy diffusivity follows from a constant turbulent Prandtl number, $\alpha_t=\nu_t/\Pran_t$ with $\Pran_t=0.9$. This is the formulation we use as a reference, with its original $\kappa=0.41$. In the formulation of \citet{ceci2025} (their \textit{wall damping IV}) the eddy viscosity and diffusivity are
\begin{equation}
\nu_t=\tilde\nu\,\frac{\chi^2}{\chi^2+c_{v1}'^2},\qquad \alpha_t=\frac{\tilde\nu}{\Pran_{t,\infty}}\,\frac{\chi_\theta^2}{\chi_\theta^2+c_{v,\theta}^2},\qquad \chi_\theta=\frac{\chi}{\Pran_{t,\infty}},\label{eq:ceci}
\end{equation}
with $c'_{v1}=7.3$, $c_{v,\theta}=10$ and $\Pran_{t,\infty}=0.843$, which yield $\nu_t\sim y^3$ and $\alpha_t\sim y^3$ at the wall, as required, instead of the $\nu_t\sim y^4$ of the standard $f_{v1}$ \citep{ceci2025}. In the logarithmic layer $\tilde\nu\simeq\kappa u_\tau y$, so that $\chi\simeq\kappa\ycp$ and $\chi_\theta\simeq\kappa_\theta\ycp$ with $\kappa_\theta=\kappa/\Pran_{t,\infty}=0.459$. Equation \eqref{eq:ceci} is then the transport-equation counterpart of the algebraic model \eqref{eq:eddy}, since $\nu_t/\bar\nu=\chi^3/(\chi^2+c'^2_{v1})$ and $\alpha_t/\bar\nu=\chi_\theta^3/(\chi_\theta^2+c_{v,\theta}^2)$. This suggests introducing the property corrections in the same way,
\begin{equation}
\nu_t=\tilde\nu\,\frac{\chi^2}{\chi^2+c_{v1}'^2+\phi},\qquad \alpha_t=\frac{\tilde\nu}{\Pran_{t,\infty}}\,\frac{\chi_\theta^2}{\chi_\theta^2+c_{v,\theta}^2+\beta},\label{eq:sawater}
\end{equation}
with $\phi$ and $\beta$ given by \eqref{eq:law}, without any recalibration. The modified damping function also enters $\tilde S$ through $f_{v2}=1-\chi/(1+\chi f_{v1})$. The turbulent Prandtl number of the model,
\begin{equation}
\Pran_t=\frac{\nu_t}{\alpha_t}=\Pran_{t,\infty}\,\frac{\chi^2+\Pran_{t,\infty}^2(c_{v,\theta}^2+\beta)}{\chi^2+c'^2_{v1}+\phi},
\end{equation}
tends to $\Pran_{t,\infty}$ away from the wall and to $\Pran_{t,\infty}^3(c_{v,\theta}^2+\beta)/(c'^2_{v1}+\phi)$ at the wall, so that the two corrections act on the near-wall turbulent Prandtl number. The fluctuating-property terms and the difference between $\overline{\mu(T)}$ and $\mu(\bar T)$ are neglected, as in \S\,\ref{sec:transf}. The equations are discretised with second-order finite volumes on 1024 cells stretched towards the walls, and solved at the $\Rey_b$, $T_w$ and $\TmTw$ of each DNS. Halving the number of cells changes the coefficients by less than $0.1\,\%$.

\begin{figure}
\centerline{\includegraphics[width=\textwidth]{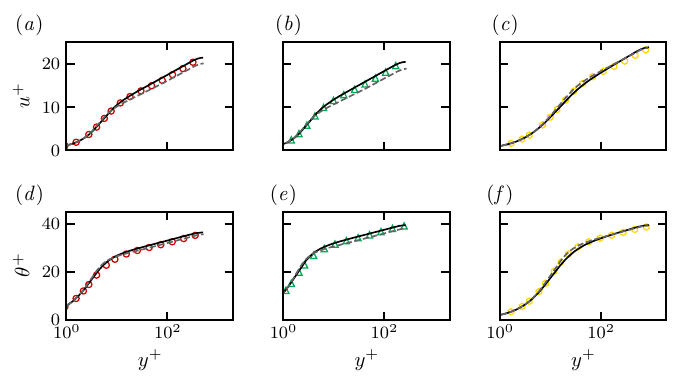}}
\caption{Mean (\textit{a--c}) velocity and (\textit{d--f}) temperature profiles in wall units for flow cases (\textit{a,d}) H12-B, (\textit{b,e}) H13 and (\textit{c,f}) H083. The symbols are the DNS data, the solid black lines the SA model with the formulation of \citet{ceci2025} and the corrections \eqref{eq:sawater}, and the dashed grey lines the standard SA model with $\Pran_t=0.9$. \jfmnb{https://cocalc.ai/share/modesti-ceci-water-channel/Figure-12/figure12.ipynb}}
\label{fig:saprof}
\end{figure}

\begin{figure}
\centerline{\includegraphics[width=\textwidth]{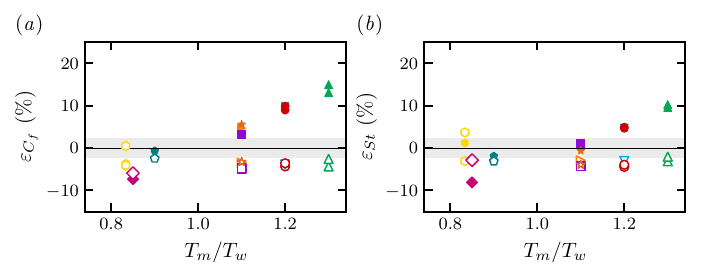}}
\caption{Percentage error of (\textit{a}) the friction coefficient and (\textit{b}) the Stanton number predicted by the SA model with respect to the DNS, as a function of $\TmTw$, for the present formulation \eqref{eq:sawater} (open symbols) and the standard SA model with $\Pran_t=0.9$ (filled symbols). The grey band denotes $\pm2\,\%$. Colours and symbols as in figure~\ref{fig:specvar}. \jfmnb{https://cocalc.ai/share/modesti-ceci-water-channel/Figure-13/figure13.ipynb}}
\label{fig:saerr}
\end{figure}

\begin{table}
\begin{center}
\def~{\hphantom{0}}
\begin{tabular}{lrrrrrr}
\toprule
 & \multicolumn{2}{c}{$\TmTw>1$} & \multicolumn{2}{c}{$\TmTw<1$} & \multicolumn{2}{c}{all}\\
\cmidrule(lr){2-3}\cmidrule(lr){4-5}\cmidrule(lr){6-7}
SA model & $C_f$ & $St$ & $C_f$ & $St$ & $C_f$ & $St$\\
\midrule
standard (I) & 8.7 & 4.8 & 4.0 & 3.6 & 7.2 & 4.4 \\
IV, $\phi=\beta=0$ & 4.6 & 4.2 & 2.8 & 4.0 & 4.0 & 4.1 \\
IV, $\phi$, $\beta$ of \eqref{eq:law} & 3.8 & 3.5 & 3.2 & 3.2 & 3.6 & 3.4 \\
\bottomrule
\end{tabular}
\caption{Mean absolute percentage error of the friction coefficient and Stanton number predicted by the SA model with respect to the DNS, for wall cooling ($\TmTw>1$, eight cases), wall heating ($\TmTw<1$, four cases) and all cases. The standard model (I) uses the damping function $\chi^3/(\chi^3+7.1^3)$, $\kappa=0.41$ and $\Pran_t=0.9$. Model IV is the formulation \eqref{eq:ceci} of \citet{ceci2025}, without and with the corrections \eqref{eq:sawater}.}
\label{tab:sa}
\end{center}
\end{table}

The mean profiles predicted by the SA model are compared with the DNS in figure~\ref{fig:saprof}, and the errors in the coefficients are summarised in figure~\ref{fig:saerr} and table~\ref{tab:sa}. With the standard model the velocity in the outer layer is too low for wall cooling, which corresponds to friction errors of $+3$ to $+6\,\%$ at $\TmTw=1.1$ and up to $+15\,\%$ at $\TmTw=1.3$. The error grows with the property contrast, as in the transformation framework without corrections. The formulation of \citet{ceci2025} alone (IV with $\phi=\beta=0$) has a smaller but still systematic friction error for wall cooling ($+5$ to $+9\,\%$ at $\TmTw\ge1.2$), and its heat transfer error for wall heating reaches $+10\,\%$ for L083. With the corrections \eqref{eq:sawater} the profiles of velocity and temperature follow the DNS closely (figure~\ref{fig:saprof}), the mean friction error for wall cooling is less than half that of the standard model, and the errors no longer depend on the property contrast (figure~\ref{fig:saerr}). $C_f$ and $St$ are predicted within $-6$ to $+1\,\%$ and $-5$ to $+4\,\%$ for all cases. For the eleven cases with $\Retcp\ge289$ the remaining error is a nearly uniform under-prediction of both coefficients, by $2$--$6\,\%$, which is not related to the property variations. For the mild ratios $\TmTw=0.9$--$1.1$ the uncorrected formulation IV is in fact already within about $\pm2\,\%$, and the corrections over-compensate. Part of this bias is the baseline error of the model at constant properties, where strategy IV under-predicts $C_f$ by about $2\,\%$ at $\Rey_\tau=5200$ with respect to \citet{leemoser2015}. Another part arises because $\phi$ and $\beta$ were calibrated for the algebraic model, in which the corrections act on $\kappa\ycp$ directly, whereas in the SA model they also modify the production through $f_{v2}$ and hence $\tilde\nu$ itself. The heat transfer predictions are also sensitive to $\Pran_{t,\infty}$, as changing it from $0.843$ to $0.8$ or $0.9$ changes $St$ by about $+5\,\%$ and $-6\,\%$, respectively, and $C_f$ by about $1\,\%$ at most. Considering that the standard SA model with constant turbulent Prandtl number errs by up to $15\,\%$ in friction and $10\,\%$ in heat transfer, and that the engineering correlations err by up to $26\,\%$, the corrected model offers a uniform accuracy of a few per cent across the whole range of conditions at no additional cost.

%=====================================================================================================
\section{Conclusions}\label{sec:conclusions}

We have studied turbulent forced convection of liquid water in a plane channel by DNS, with viscosity and conductivity depending on temperature, over a range of wall heating and cooling conditions in which the viscosity varies by up to a factor of five between the wall and the mixed mean temperature and the Prandtl number varies across the channel. The database shows that the property variations alter the friction and heat transfer coefficients by tens of per cent, well beyond the tolerance of thermal design, and that the correlations in use since the 1930s err by amounts comparable to the property effects themselves, with an error that grows with the viscosity contrast.

The physical picture that emerges is simple. The property variations act in the wall layer, where the local viscous length, rather than the wall one, sets the size of the near-wall eddies and the spacing of the velocity and temperature streaks. The friction Reynolds number based on the wall viscosity therefore misrepresents the scale separation, and the equivalent constant-property Reynolds number $\Retcp$, which measures the channel half-width in units of the local viscous length at the centreline, is the appropriate parameter for comparing flow cases and for sizing meshes. Away from the wall, the constant density makes the distributions of shear stress and heat flux independent of the properties, and the untransformed velocity and temperature defects follow those of constant-property flow. The whole effect of the property variations on friction and heat transfer is thus confined to a thin layer at the wall, which is what a predictive model has to describe.

This observation is what makes the transformation approach work. The integral transformations derived for air carry over to liquids with minor changes, which account for the decoupling of conductivity from viscosity and for the variable Prandtl number. That the structure of the model holds for air and water, two fluids with opposite dependence of the viscosity on temperature, with and without density variations, indicates that it rests on the mechanics of the wall layer rather than on the properties of a specific fluid. The Spalart--Allmaras model deserves credit in this respect. Its damping function depends on $\chi=\tilde\nu/\bar\nu$, the ratio of the working variable to the local viscosity, so that the wall layer is scaled semi-locally by construction, as noted by \citet{hasan2025rans} for compressible flows, and the model follows the change of the viscous length across the layer without any correction. What it misses is the modification of the near-wall turbulence beyond this change of scale, which the corrections $\phi$ and $\beta$ supply. They transfer to the model without recalibration and remove the dependence of its errors on the property contrast, so that the approach can be implemented in the RANS solvers used in engineering practice, where variable-property heat transfer is often still handled through the correlations discussed above.

A broader remark concerns the nature of the problem. Heat transfer with variable properties is not universal, unlike the passive scalar, whose statistics depend on the Reynolds and Prandtl numbers alone. The property laws introduce dimensional parameters, the wall temperature first of all, so that two flows with the same Reynolds number, Prandtl number and temperature ratio differ if their wall temperatures differ, as the cases at $\TmTw=1.1$ and $1.2$ show. This should not discourage its study. The underlying equations hold regardless, and the transformations use them to connect the universal passive-scalar problem to this more complex one, so that the constant-property statistics remain the reference from which the variable-property predictions are built, whatever the fluid and the operating conditions.

Several questions remain open. The simulations neglect buoyancy, which in mixed convection alters the wall layer in a way that depends on the orientation of the heat flux with respect to gravity, and the walls are smooth, whereas the surfaces of heat exchangers are often rough, with a roughness sublayer that interacts with the variable viscosity in a way that has not been explored. The heating is uniform and symmetric, while heat exchangers are typically heated on one side, with a non-uniform heat flux along the wall, so that the property variations differ across the section and along the flow. Finally, the extension of the transformations to pipes and ducts, and to boundary layers, whose outer layer differs from that of the channel, remains to be done.

\begin{bmhead}[Acknowledgements.]
We acknowledge the EuroHPC Joint Undertaking for awarding us access to LEONARDO, hosted by CINECA, Italy, through the project \mbox{EUHPC\_E06\_110}, and to Arrhenius, hosted by the National Academic Infrastructure for Supercomputing in Sweden (NAISS), through the project \mbox{EHPC-REG-2026r01-287}.
\end{bmhead}

\begin{bmhead}[Funding.]
This work was supported by the Italian Ministry of University and Research (MUR) through the Fondo Italiano per la Scienza (FIS\,2) Starting Grant project HEATFORCE, ``predicting HEAT transfer by FORCEd convection in variable properties fluids'', grant no.~FIS-2023-00346, CUP D53C25002050001.
\end{bmhead}

\begin{bmhead}[Use of artificial intelligence.]
The authors used Claude Code (Anthropic, model Claude Fable 5.1, claude.ai/code) between October and November 2026 to draft and revise parts of the text, to write the Python scripts that produce the figures and tables from the simulation statistics, and to prepare the JFM Notebooks. All text, code and results were checked and edited by the authors, who take full responsibility for the content. No figure was generated or modified by AI beyond the plotting of the simulation data.
\end{bmhead}

\begin{bmhead}[Declaration of interests.]
The authors report no conflict of interest.
\end{bmhead}

\begin{bmhead}[Data availability statement.]
The data and the code that generate every figure of the paper are available as JFM Notebooks, linked from the figure captions and provided as supplementary material. Each notebook contains the time-averaged DNS statistics, spectra and instantaneous planes shown in the figure, the constant-property reference profiles, the predictions of the transformations and the solutions of the Spalart--Allmaras model, together with the parameters of all flow cases. The property laws and the inputs of the simulations are given in \S\,\ref{sec:method}.
\end{bmhead}

\bibliographystyle{jfm}
\bibliography{refs}

\end{document}